\documentstyle[aps,prl,multicol,epsf]{revtex}
\begin{document}

\title{\bf Phase Transition in a Random Fragmentation Problem with Applications to Computer 
Science}
\author{David S. Dean and Satya N. Majumdar}
\address{CNRS, IRSAMC, Laboratoire de Physique Quantique,
Universit\'e Paul Sabatier, 31062 Toulouse, France} 

\maketitle

\begin{abstract}{
We study a fragmentation problem where an initial object of size 
$x$ is broken into
$m$ random pieces provided $x>x_0$ where $x_0$ is an atomic cut-off. Subsequently the 
fragmentation
process continues for each of those daughter
pieces whose sizes are bigger than $x_0$. The process stops when
all the fragments have sizes 
smaller than $x_0$. We show that the fluctuation of the total number of 
splitting
events, characterized by the variance, generically undergoes a nontrivial
phase transition as one tunes the branching number $m$ through 
a critical value $m=m_c$.
For $m<m_c$, the fluctuations are Gaussian where as for $m>m_c$ they are 
anomalously
large and non-Gaussian. We apply this general result to analyze two 
different search algorithms
in computer science.}

\noindent

\date{22 April 2002}
\end{abstract}

\begin{multicols}{2}
Fragmentation is a widely studied phenomena\cite{review} 
with applications ranging from conventional fracture of solids\cite{LW}
and collision induced fragmentation in atomic nuclei/aggregates\cite{nuc}
to seemingly unrelated fields such as disordered systems\cite{disorder}
and geology\cite{geology}. In this paper we consider a 
problem where an object of initial size (or length) $x$ is first broken into $m$ random pieces of 
sizes
$x_i=r_i x$ with $\sum_{i=1}^m r_i=1$ provided the initial size $x>x_0$
where $x_0$ is a fixed `atomic' threshold. At the next stage, each of those $m$ pieces
with sizes bigger than $x_0$ is further broken into $m$ random pieces and so on.
Clearly the process stops after a finite number of fragmentation or splitting events
when the sizes of all the pieces become less than $x_0$. This problem and its close cousins
have already appeared in numerous contexts including the energy cascades in 
turbulence\cite{turb}, rupture processes in earthquakes\cite{earthq}, stock market 
crashes\cite{finance}, binary search algorithms\cite{Devroye,krma,makr}, stochastic 
fragmentation\cite{KBG} and DNA segmentation algorithms\cite{DNA}. It therefore comes
as somewhat of a surprise that there is a nontrivial phase transition in this
problem as one tunes the branching number $m$ through a critical value $m=m_c$.

In this Letter we study analytically the statistics of the total number of fragmentation 
events $n(x)$ up till the end of the process as a function of the initial size $x$. We show that, 
while 
the average number
of events $\mu(x)$ always grows linearly with $x$ for large $x$, the 
asymptotic
behavior of the variance $\nu(x)$, characterizing the fluctuations,
undergoes a phase transition at a critical value $m=m_c$,
\begin{equation}
\nu(x)\sim\cases
                   {x                 &$m<m_c$,\cr
                    x^{2\theta}     &$m\geq m_c$\cr}.
\label{asymp}
\end{equation} 
The exponent $\theta$ is nontrivial and increases monotonically with $m$ for $m\geq m_c$ starting
at $\theta(m=m_c)=1/2$ and the amplitude of the leading $x^{2\theta}$ term
has 
log-periodic oscillations for $m\geq m_c$. This signals
 unusually large fluctuations in $n(x)$
for $m>m_c$.
The full distribution of $n(x)$ also changes from being Gaussian for $m<m_c$ to non-Gaussian
for $m>m_c$.
This
phase transition is rather generic for any fragmentation problem with an `atomic'
threshold. However the critical value $m_c$ and the exponent $\theta$ are nonuniversal
and depend on the distribution function of the random fractions $r_i$'s. In this Letter
we establish this generic phase transition and then calculate explicitly $m_c$
and $\theta$ for two special cases with direct applications in computer science. 
 
In this fragmentation problem with a fixed lower 
cut-off $x_0$, one first breaks the initial piece of length $x$ provided $x>x_0$
into $m$ pieces of sizes $x_i=r_i x$. The sizes of each of these `daughters'
are then examined. Only those pieces whose sizes exceed $x_0$ are
considered `active' and those with sizes less than $x_0$ are considered `frozen'.
Each of the active pieces is then subsequently broken into $m$ pieces and so on. 
The fractions
$r_i$'s characterizing a splitting event are considered to be
independent from one event to another but are drawn each time from the
same joint distribution function $\eta_m(r_1,r_2, \ldots r_m)$. As the splitting
process conserves the total size, the fractions $r_i$'s satisfy the constraint
$\sum_{i=1}^m r_i=1$. In addition, we consider the splitting process to be isotropic, i.e.,
all the $m$ daughters resulting from a splitting event are statistically equivalent. This
indicates that the marginal distribution of
any one of the $r_i$'s is independent of $i$ and is given by,
\begin{equation}
\eta_1(r) = \int\ \eta_m(r,r_2,\cdots r_m) \prod_{i = 2} ^m dr_i .
\end{equation}
We will henceforth denote the average over the whole history of the splitting
procedure (till the end of the process) as ${\overline{\cdots}}$ and 
the average over the $r_i$'s
associated with a single splitting event as $\langle \cdot \rangle$. 
The conservation law along with the isotropy implies that
$\langle r \rangle = \int \eta_1(r) r \ dr = 1/m$. 

Clearly the total number of splitting events $n(x)=0$ if $x<x_0$. On the other hand if $x>x_0$ 
there will be at least one splitting
and it is easy to write a recursion relation for $n(x)$,
\begin{equation}
n(x) = 1 + \sum_{i=1}^m n(r_i x).
\label{eqn}
\end{equation}
Using the 
isotropy of the splitting distribution and taking the 
average over Eq. (\ref{eqn}), we find
that $\mu(x) \equiv \overline{n(x)}$ satisfies the recursion  
for $x > x_0$,
\begin{equation}
\mu(x) = 1 + m \langle \mu(rx)\rangle = 1 + m
\int_{x_0/ x}^1 dr\ \eta_1(r)\mu(rx),
\label{eqnav}
\end{equation}  
where the lower limit in the above integral comes from the condition
$n(x) = 0$ for $x < x_0$. Without any loss of generality we set $x_0 = 
1$, {\em i.e.}, we measure all sizes in units of the atomic size. Since
$1\leq x < \infty$ in Eq. (\ref{eqnav}), it is 
convenient to make a change of variable $x = e^{\alpha}$ so that
$0\leq \alpha <\infty$ and write
$\mu( e^{\alpha}) = F(\alpha)$. The resulting equation for $F(\alpha)$ is 
solved by taking the Laplace transform of Eq. (\ref{eqnav}) and one
finds that ${\tilde F}(s) = \int_0^\infty d\alpha \ F(\alpha) e^{-s \alpha}$
is given by
\begin{equation}
{\tilde F}(s) = {1\over s\left[ 1 - m w(s)\right]},
\label{eqlap}
\end{equation}  
where $w(s) =\langle r^s\rangle = \int_0^1 dr\ \eta_1(r)\ r^s$. Assuming
${\tilde F}(s)$ has simple poles at $s=\lambda_k$, the Laplace transform in Eq. (\ref{eqlap})
can be inverted to obtain
$\mu(x) = a_0 + \sum_k a_k x^{\lambda_k}$ with $a_0 = 1/(1-m)$ (coming from the
pole at $s=0$) and
$a_k = -1/[m\lambda_k w'(\lambda_k)]$. From the conservation law
$\langle r\rangle =1/m$, one finds that $s=1$ is always a
pole of ${\tilde F}(s)$. Besides, since $0\le r\le 1$, the pole at $s=1$
is also the one with the largest real part and hence will dominate the 
large $x$ behavior of $\mu(x)$. Let $\lambda$ and $\lambda^*$ denote
the pair of complex conjugate poles with the next largest real part.
Then keeping only the leading corrections to the asymptotic behavior 
one finds
\begin{equation}
\mu(x) \approx a_1 x +  a_2 x^{\lambda} + a_2^* x^{\lambda^*},
\label{eqmuapprox}
\end{equation}  
where $a_1 =-1/[mw'(1)]=-1/[m\int_0^1 dr\ \eta_1(r) r\ln(r)]$.

We now turn to the variance $\nu(x)$ of the total number of splittings $n(x)$:
\begin{equation}
\nu(x) =  \overline{\left(n(x) - \mu(x)\right)^2}.
\end{equation}
By squaring Eq. (\ref{eqn}) and after some straightforward algebra we find
the recursion relation
\begin{equation}
\nu(x) = f(x) + m \int_{1/ x}^1 dr\ \eta_1(r) \nu(rx),
\end{equation}
where $f(x) = \langle \left( \sum_{i=1}^m \left[\mu(r_i x) - \langle 
\mu(rx)\rangle\right]  \right)^2 \rangle $.
Once again the change of variable $x=e^{\alpha}$ followed by a subsequent Laplace transform with 
respect to $\alpha$, ${\tilde{\nu}}(s)=\int_0^{\infty} d{\alpha} \nu(e^{\alpha})e^{-s\alpha}$
yields, 
\begin{equation}
\tilde{\nu}(s) = {\tilde {f}(s) \over  s\left[ 1 - m w(s)\right]}.
\label{nus}
\end{equation}
Using the asymptotic expression of $\mu(x)$ from Eq. (\ref{eqmuapprox}) in the expression for 
$f(x)$ one finds
that the leading term 
of $f(x)$ for large
$x$ is given by
\begin{equation}
f(x) \approx 
b_1 x^{2\lambda}+ b_2 x^{2\lambda^*} + b_3 x^{(\lambda+\lambda^*)},
\end{equation}
where the  $b_j$'s are constants. 
These behaviors indicate that ${\tilde f}(s)$ has poles at $s=2\lambda$, $s=2\lambda^*$
and $s= \lambda+\lambda^*$. Thus when $Re(\lambda)<1/2$, these poles occur
to the left of $s=1$ in the complex $s$ plane. From Eq. (\ref{nus}) it follows
that the asymptotic large $x$ behavior of $\nu(x)$ will then be controlled by the
$s=1$ pole arising from the denominator $\left[1-mw(s)\right]$ and $\nu(x)\sim x$
for large $x$. On the other hand when $Re(\lambda)>1/2$, the dominant poles
governing the large $x$ behavior are the three poles of ${\tilde f}(s)$ with
real part $2 Re(\lambda)>1$. Hence in that case, $\nu(x)\sim x^{2 \theta}$ where
$\theta=Re(\lambda)$. Note also that for $Re(\lambda)\geq 1/2$, the amplitude
of the leading term $x^{2\theta}$ in $\nu(x)$ will have 
log-periodic oscillations
due to the nonzero imaginary
parts of the poles $\lambda$ and $\lambda^*$.
This phase transition will 
always 
occur whenever 
one can tune the pole $\lambda$ continuously through the critical value $Re(\lambda)=1/2$.
In the following two examples 
we show explicitly that this can be achieved, in a natural way, by tuning
the branching number $m$.
\begin{figure}
\epsfxsize=200pt
\narrowtext\centerline{\epsfxsize\columnwidth \epsfbox{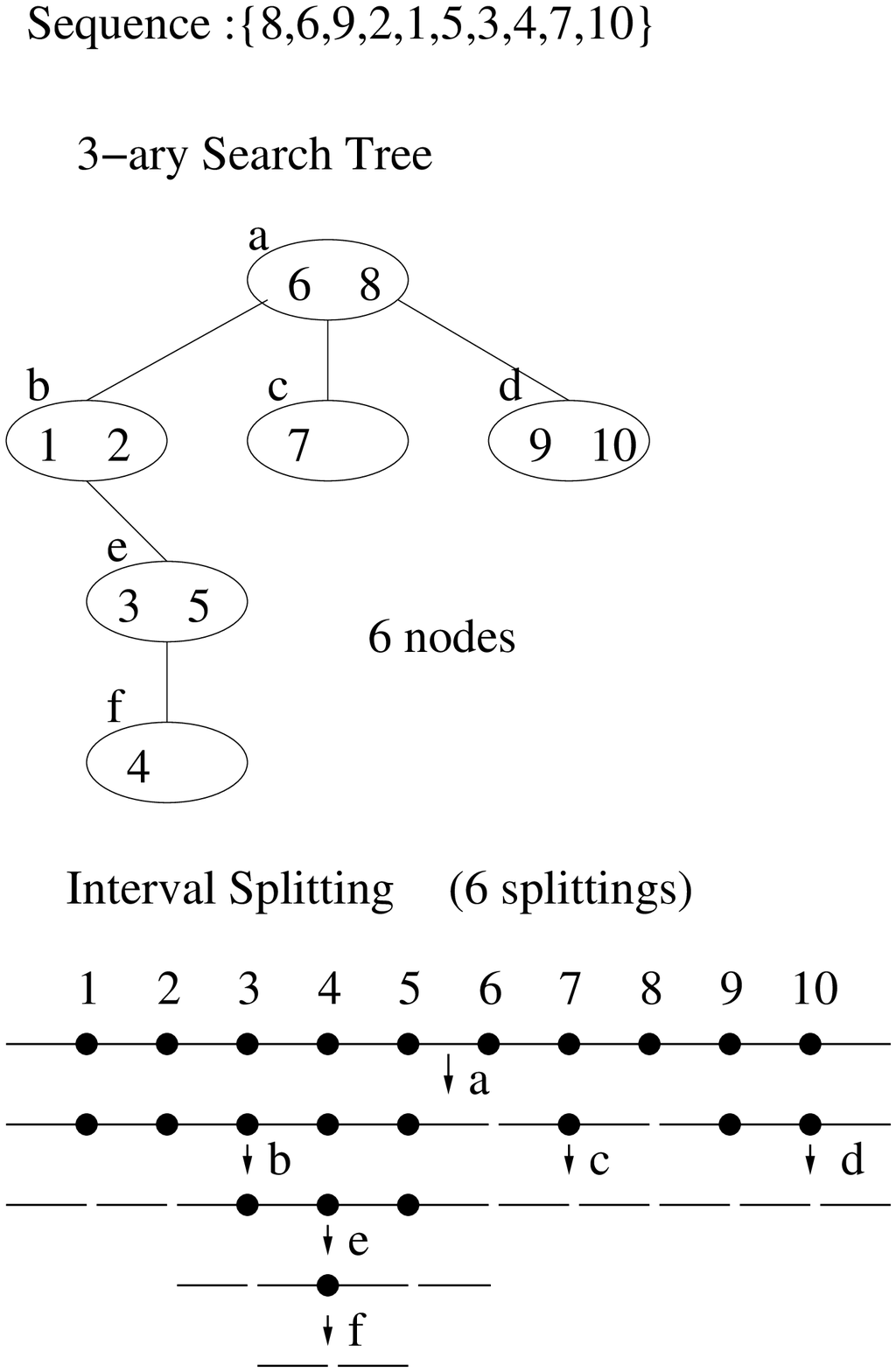}}
\caption{Top: The construction of the 3-ary tree for a sequence of numbers
between 1 and 10. Bottom: The induced random interval splitting. The nodes and 
corresponding splittings are labelled a,b,c,d,e and f. An interval can
only split if it contains a $\bullet$.}
\label{mary}
\end{figure} 
{\bf The m-ary search tree:} It is well known that one of the most efficient ways
to sort  the incoming data to a computer is to organize the data on
a tree\cite{Knuth}. Consider the sorting of an incoming data string consisting of
$N$ distinct elements labelled by the sequence $1,\ 2,\  \cdots N$.
Consider a particular random sequence of arrival of these $N$ elements (see Fig. 1).  
An m-ary search tree stores this sequence
on a growing tree structure where each node of the tree can contain at
 most $(m-1)$ data
points\cite{Devroye,makr}. A node if  filled branches into 
$m$ leaves. 
The first $(m-1)$ elements are stored in the root of the tree in an ordered
sequence $w_1<w_2 \cdots <w_{m-1}$. 
Any subsequent element must belong to   
one of $m$ sets of numbers $A_1 = [1, w_1)$,
$A_i = (w_i, w_{i+1})$ with $1\leq i \leq m-2$ and $A_{m} = (w_{m-1}, N]$.
Associated with each of these sets or intervals we associate a leaf of the 
tree leading to a new node. A new data point $w$, arriving subsequently, is sent down the leaf
corresponding to the set $A_i$ if $w\in A_i$
and is stored in a daughter node at the base of that leaf.
Once a daughter node is filled
with $m-1$ numbers it in turn gives rise
to $m$ new leaves and so on. An example for a 3-ary search tree
with $N = 10$ is shown in Fig. (\ref{mary}). 

Each sequence of the incoming data will give rise to a different $m$-ary tree
configuration. If the incoming data is random, all the trees occur with equal
probability. It is easy to see that the total number of occupied nodes 
$M$ (each
containing at least one element) is a random variable
as it varies from one tree configuration to another, except for $m=2$ where $M=N$. The 
statistics 
of $M$ was recently studied by computer scientists using rather involved combinatorial
analysis and it was found that while ${\overline {M}}\sim N$ for large $N$,
the variance $\nu \sim N $ for $m<26$ and as $\sim N^{2\theta}$ for $m>26$\cite{compsci}.
We show below that this strange result is just a special case of the general
phase transition in the fragmentation problem discussed here.

The construction of the $m$-ary search tree can be mapped exactly onto the 
splitting
of the interval $[1,N]$ \cite{Devroye,makr}.  
It is easy to see that the incoming  
elements $w_1, \ w_2 \cdots w_{m-1}$ split the initial interval into $m$
parts $A_i$. If all the $N!$ possible sequences arrive with equal probability
then the points $w_1, \ w_2 \cdots w_{m-1}$ 
are distributed uniformly on $[1,N]$ (these are the numbers 
stored in the first node). We split the interval $[1,N]$ into 
$m$ subintervals corresponding to the $A_i$'s introduced above. 
If a subinterval $A_i$ is empty (i.e. has no $\bullet$ in Fig. (2)) 
then no data points can go down 
the 
corresponding leaf and hence such an interval (of length $<2$) will not split
any further. If the subset $A_i$ contains only one $\bullet$, 
the arrival of the corresponding single data point 
still splits the interval into $m$ parts 
(some of the intervals so created may be of length 0).   
This corresponds to the atomic threshold $x_0=2$ in our general
problem and $x=N/2$ corresponds to the initial size in units of 
the atomic size. 

The crucial point is
that the number of occupied nodes $M$ in the $m$-ary 
search tree is identical to the number of splittings $n(x=N/2)$ 
in this fragmentation problem. For large $N$ one can pass to a continuum 
limit  and use the known marginal probability density function
$\eta_1(r)=(m-1) (1-r)^{m-2}$\cite{krma,makr} for the continuum 
interval splitting
problem in our general formula. We get ${\overline {M}}=\mu(x=N/2) \approx {a_1 N}/2$ for 
large 
$N$ where $a_1= 1/[{\sum_{k=2}^m} 1/k]$. Also $w(s)=\langle r^s\rangle = (m-1)B(s+1,m-1)$
where $B(m,n)$ is the standard Beta function. Therefore the poles of ${\tilde F}(s)$
in Eq. (\ref{eqlap}) occur at the roots of the equation
$m(m-1)B(s+1, m-1)=1$. It is easy to check
using Mathematica that one can arrive at the critical condition 
$Re(\lambda) = 1/2$ by tuning $m$
through the value $m=m_c\approx 26.0461\cdots$. Therefore, from our general theory, we find
that the variance $\nu \sim N$ for $m<m_c$ and  $\nu\sim N^{2\theta}$
for $m>m_c$. The exponent $\theta=Re(\lambda)$ where $\lambda$ 
is the root of
$m(m-1)B(s+1, m-1)=1$ that is closest (to the left) to $s=1$ when $m>m_c$.
\begin{figure}
\narrowtext
\epsfxsize=0.8\hsize
\epsfbox{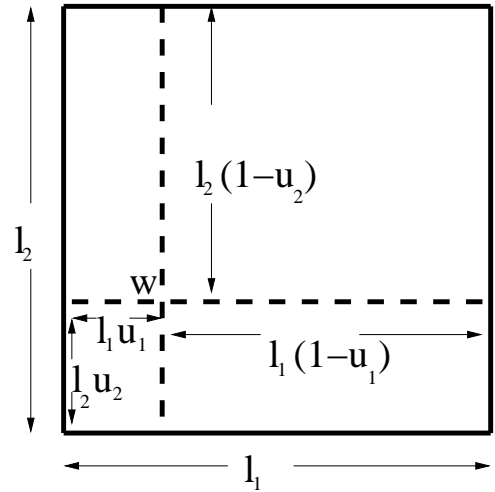}
\caption{An example of the splitting of a rectangle into four daughter
rectangles about the point ${\bf w} = [l_1 u_1,l_2 u_2]$. The
same process is continued on each daughter until the area of the
daughter becomes less than $x_0$.}
\label{figcube}
\end{figure}
{\bf Cuboid Splitting:} In the previous problem we have considered the sorting
of a data string where each element is a scalar. A natural generalization
is when each element is a $D$-dimensional vector ${\bf w}$ whose $k$-th component 
$w_k\in [0,l_k]$ for $1\leq k\leq D$. The first element of the data string
is then assigned to the point ${\bf w}$ 
in the cuboid of edge lengths $l_k$. 
If the first element is random, then its components $w_k=u_k l_k$, where
the $u_k$ are independent random variables uniformly distributed on $[0,1]$.
Once this first element is stored, it
splits the original cuboid into
$2^D$ sub-cuboids obtained by drawing $D$ lines perpendicular to each of the
faces of the cuboid (see Fig. 2). When the 
second vector 
arrives, one compares
its components with that of the first element and places it in one of the $2^D$
sub-cuboids (and thereby splits that sub-cuboid) and the process continues.  
After each splitting event, the dimensions of the $m=2^D$ new sub-cuboids
can be
represented by $l'_k(\sigma) = l_k u_k (1+ \sigma_k)/2 +  l_k (1-u_k)   
 (1- \sigma_k)/2$, where the $\sigma_k$ are Ising spins.
An example with $D=2$ is shown in Fig(\ref{figcube}).

The volume of any of the sub-cuboids upon splitting the cuboid of volume $x$
is $x' = xr(\sigma)$, where 
$r(\sigma) =  \prod_{k=1}^D\left(u_k (1+ \sigma_k)/2 + (1-u_k) (1- \sigma_k)
/2\right)$. Hence in this problem one has $m = 2^D$ and the 
marginal distribution of a given $r(\sigma)$ can be shown to be
\begin{equation}
\eta_1(r) = {[-\ln(r)]^{D-1}\over (D-1)!}, \ \ \ 0\leq r\leq 1.
\end{equation}
From Eq.(\ref{eqlap}) with this marginal distribution one finds that
\begin{equation}
{\tilde F}(s) = {1\over {s\left( 1 - {2^D\over (s+1)^D} \right)}}.
\label{cuboid}
\end{equation}
One then finds $\mu(x) \approx 2x/D$ for large $x$. The function ${\tilde F}(s)$
has a total of $(D+1)$ poles: one at $s=0$ and the others at
$s=-1+2 e^{2\pi i n/D}$ with $n=0,1,\ldots (D-1)$. The 
poles closest to the left of $s=1$ are the complex conjugate pair
$\lambda= -1 +2 e^{2\pi i/D}$ and $\lambda^*=-1+2 e^{-2\pi i/D}$.
Thus $Re(\lambda)= -1+2 \cos( 2\pi/D)$. From our general theory,
it follows that by tuning $m$ or equivalently $D$, it is possible
to encounter the critical point $Re(\lambda)=1/2$ at $D=D_c=
 \pi/ \sin^{-1}(1/2\sqrt{2})
= 8.69\ldots$. Hence the variance $\nu(x) \sim x$ for $D<D_c$,
and for $D\geq D_c$, $\nu(x) \sim x^{2\theta}$ where $\theta=Re(\lambda)=2\cos(2\pi /D) -1$.

\begin{figure}
\narrowtext\centerline{\epsfxsize\columnwidth \epsfbox{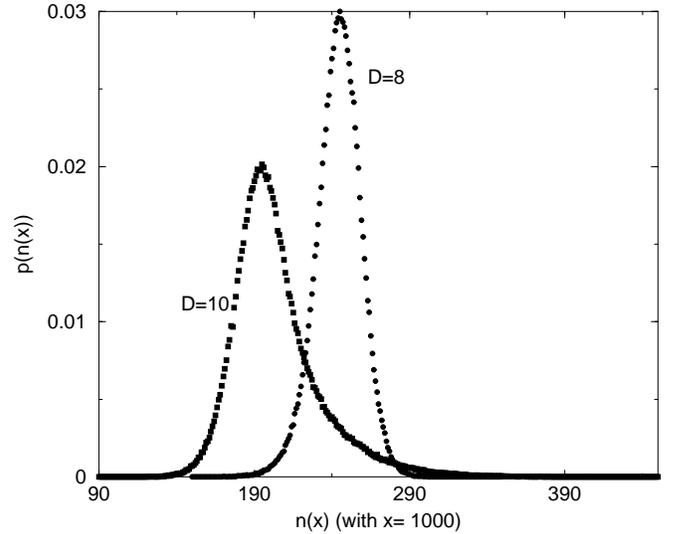}}
\caption{The distribution $p(n(x))$ of the number of splittings of a cuboid
of original volume $x=1000$ for $D = 8$ (filled circles) and for 
$D=10$ (filled squares). The distribution is Gaussian for $D=8$, but
has a non-Gaussian skewness for $D=10$. The histogram was 
formed by numerically splitting $5\times 10^5$ samples in each case.}
\label{hist}
\end{figure}

We have verified the above predictions by numerically  
carrying out the splitting procedure on a large number of samples
with atomic cut-off $x_0 = 1$. The analytical predictions
for the mean and the variance are well verified 
though an accurate measurement of the 
exponent $\theta$ is difficult due to statistical fluctuations and
finite size corrections. We have also measured the  
histogram of the number of 
splittings. For $D< D_c$ this distribution is Gaussian, 
however for $D> D_c$ the distribution becomes skewed towards large values
of $n(x)$ having an anomalous tail. Shown in Fig. (\ref{hist}) is 
the distribution measured for $D=8$ and $D= 10$. The difference
is clearly visible. The non-Gaussian behavior is also visible for the case 
$D=9$ but less pronounced as $Re(\lambda)$ is quite close to $1/2$. 
While we can rigorously prove that the distribution is indeed Gaussian
in the sub-critical regime, we have not been able to calculate the full
distribution in the supercritical regime.
Qualitatively it is clear however that as $D$
increases the volume of the cuboid becomes more concentrated about its 
surface and hence the splitting point ${\bf w}$ for large $D$ is generically 
closer to the surfaces. This means that the splitting procedure will
tend to cut the cuboid into more unequal pieces than at lower dimensions, 
a mixture of {\em blocks} of larger volume and {\em slices} 
of smaller volume. It is thus the blocks which are sliced rather 
than split in their middle which contribute to the long tail in the 
distribution of $n(x)$. 

In conclusion we have shown that a fragmentation process with an atomic threshold
can undergo a nontrivial phase transition in the fluctuations of the number
of splittings at a critical value of the branching number $m$. The calculation
of the full probability distribution of the number of splittings remains
a challenging unsolved problem. We have
provided applications of our general results in two computer science problems. 
The mechanism of this transition is remarkably simple and therefore
one expects it to be rather generic with broad applications
since many random processes can be mapped to the type of
fragmentation model considered here.


\end{multicols}
\end{document}